# Resonant response in mechanically tunable metasurface based on crossed metallic gratings with controllable crossing angle

V. Yachin,[1,a)] L. Ivzhenko,[2] S. Polevoy,[2] and S. Tarapov[2,3,4]

[1]*Theoretical Radio Physics Dept., IRA NASU, 4, Mystetstv Str., Kharkiv, 61002, Ukraine*

[2]*Radiospectroscopy Dept., IRE NASU, 12, Ak. Proskura Str., Kharkiv, 61085, Ukraine*

[3]*Karazin Kharkiv National University, 4, Svobody Sq., Kharkiv, 61022, Ukraine*

[4]*National University of Radio Electronics, 14, Nauka Ave., Kharkiv, 61166, Ukraine*

We report on a new class of mechanically tunable planar metamaterials comprising resonating units formed by crossed metallic strip gratings. We observe a resonant response in transmission spectra of a linearly polarized wave passing through the system of crossed gratings. Each grating consists of an array of parallel metallic strips located on the top of a dielectric substrate. It is revealed that the resonant position appears to be dependent on the angle of gratings crossing. It is found out both theoretically and experimentally that the resonant shift on the frequency scale appears as a result of increasing in the length of the resonating portion of the parallelogram periodic cell formed by the crossed metallic strips with decreasing crossing angle and the proposed design can be used in new types of planar metamaterials and filters.

In recent decades, a huge number of publications appeared which are related to the study of the optical properties of metamaterials. Metamaterials are composites possessing characteristics that cannot be found in nature (see, for instance, [1,2] and references therein). In such artificial systems the unit cell serves as an atom or molecule in conventional natural materials, whereas it can be adjustable through varying cell's geometry and constituents. Those of them that are produced by planar technology remains among the most promising for applications. They are also known as metasurfaces [3, 4].

Typically metasurfaces manifest a resonant response due to excitation of the magnetic and/or electric modes. This resonant response of metasurfaces is a cornerstone for achieving exotic behaviors which are dependent on the composition and structure of the structures' unit cell [5, 6].

___________________________

[a)] Electronic mail: yachin@rian.kharkov.ua.

Thus, there are various ways to change the resonant response. For instance, varying the meta-atom's scale or changing the relative orientation and proportions of its components one can achieve high quality factor of resonances or adjust their position on the frequency scale [7-9]. Mechanical symmetry restructuring of meta-atom can be specified as one of the possible mechanisms for resonance tuning, for example, mechanical stretching of metamaterials which is already realized on elastic substrates [10-12]. Nevertheless, in order to change the resonant response of the structure, we propose to modify not only the design of its unit cell but also the symmetry of the metasurface lattice itself. For planar 2D crystal lattice we know five distinct lattice types [13]. There are an oblique lattice or general lattice and four special lattices. Recently theoretically it was predicted that high-Q Fano transmission/reflection resonances can dramatically change their position on the frequency scale for the free standing grating of crossed metallic strips when the grating skew angle is changed [14]. The location of the resonant peak depends on the crossing angle $\beta$ and appears to be near the wavelength which correspond to $\lambda_0 \sin\beta = a$ (see Fig. 1(a)), where $\lambda_0$ is the wavelength of the incident wave.

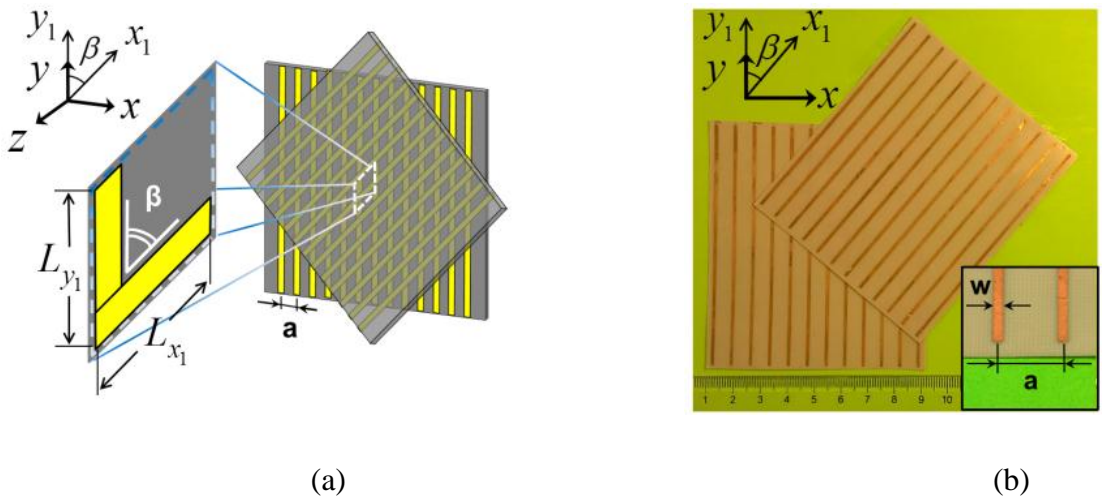

(a)                          (b)

FIG. 1. (Color online) (a) Planar metamaterial with an oblique lattice. The dashed box indicates an elementary translation cell of the structure; (b) the pattern of two fabricated metallic strip gratings placed on a dielectric substrate which forms a metamaterial under study

In this letter, we report on the observation of extremely narrow transmission resonant response in planar metamaterials which can be referred as *metasurfaces with an oblique lattice*. The resonances appear as a result of changes in the lattice symmetry and it corresponds to the field pattern in the vicinity of the perimeter of a conductive cell of the structure. The appearance of the narrow resonances is attributed to the excitation of well-known ring–like resonator mode resonances. In our experiments we use a metamaterial in order to confirm the resonances calculated in [14]. It consists of two identical slabs made of high frequency laminate Taconic TLC-30-0200-CH/CH ($\varepsilon'$=3.0, tg$\delta$=0.0024). The overall sizes of the slabs used here are 90 x 90 mm$^2$. The slab composed of a dielectric substrate with periodic metallic strips are etched from $\Delta h_i$=18 μm copper cladding covering substrate of 0.5 mm thickness. Thus, each grating consists of 12 strips with width *w*= 1 mm and length of 90 mm. The period *a* between the strips is 7 mm. The slabs are arranged one under another so that the strips appear to be crossed at a certain angle with close proximity to each other (see Fig. 1(b)). These two slabs are fixed by the plates of foam plastic ($\varepsilon'$=1.15, tg$\delta$=0.0024) for their closer fit. They are located on both sides of the planar composite and have negligible effect on the frequency response of the transmission coefficient. The thickness of the foam plastic is 20 mm. We study transmission spectra of the sample at normal wave incidence in the spectral range of 22-40 GHz using the vector network analyzer (N5230A) and two broadband horn antennas. To measure the transmission spectra the comprehensive experimental technique based on the scanning module was designed and described by authors in [15, 16]. In this structure, only the fundamental wave propagates at the frequencies lower than 40 GHz, the other diffraction orders are damped.

We consider a double layered crossed-grating formed by the arrays of parallel metallic strips of rectangular cross section $\Delta h_i \times w_i$ (*i*=3,4 are the numbers of layers of the sandwiched metallic crossed gratings (see inset in Fig. 2) with the same pitch *a* between the rods where the metal strips of both arrays are in contact with each other. The crossing angle between two arrays is *β* (see Fig. 1(a)).

Generally, the transmission properties of metasurfaces strongly depend on the polarization state of the incident electromagnetic wave since the unit cell of a structure has no π/2 rotation symmetry like a square or a circle [17]. Nevertheless, in our case a metasurface exhibits the nearly polarization independent behavior of the transmission resonant response. Firstly, we consider an incident wave $\mathbf{E_0}$ linearly polarized in the *y*-direction, when vector $\mathbf{E_0}$ is parallel to the strips of one of the gratings, see Fig. 1. For the orthogonal polarizations viz. $\mathbf{E_0}$ linearly polarized in the *x* direction, the structure with rhomb-shaped elementary unit cell does not show significant different spectral features comparatively to those of y-polarized.

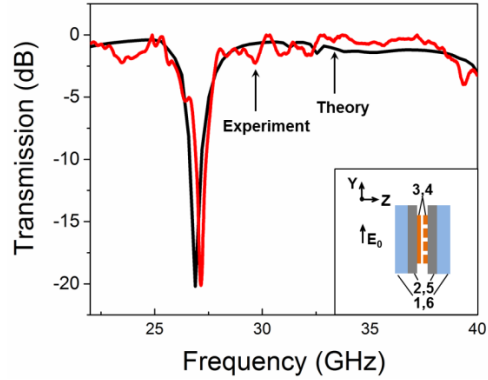

FIG. 2. (Color Online) Transmission spectra of the metasurface with crossed metallic gratings for *y*- polarization: red line—experiment, black line—theory (method of functionals). The inset shows schematic picture of the structure: (2,5)- two dielectric slabs with (3,4) metallic strips which form planar metamaterial, (1,6)- plates of foam plastic.

Measured transmission coefficient for normally incident plane wave whose E-vector is directed along the *y*-axis, is presented in Fig. 2. The curves show the measured transmission spectrum and theoretical calculation obtained by the method of integral functionals [14, 18], respectively, for planar metamaterial based on crossed metallic gratings, when crossing angle is $45^0$. One can observe a very narrow resonant transmission dip near 27 GHz, where transmission drops to about -20 dB. The observed resonant peak is in a good agreement with the theoretical prediction. The *Q*-factor of the transmission resonant response for the gratings at crossing angle $\beta = 45°$ for measured data in Fig. 2 is about 40 (we define the *Q*-factor as the ratio between the resonance frequency and the full width at half minimum of transmission). According to our estimations, *Q*-factors of the resonant peaks observed in the composite for all considered crossing angles are in the range of 40-60. Such Q-factors of resonances are much greater than those ones for planar metamaterials based on split ring resonator [19].

To predict the nature of the resonant response, the metamaterial were numerically simulated using the well verified method of integral functionals [14, 18]. Our study is based on the frequency domain method using the concept of a double-periodic magneto-dielectric layer. The method bases on the coupled volume integral equations for the equivalent electric and magnetic polarization currents in the periodic layer induced by the field of the incident wave.

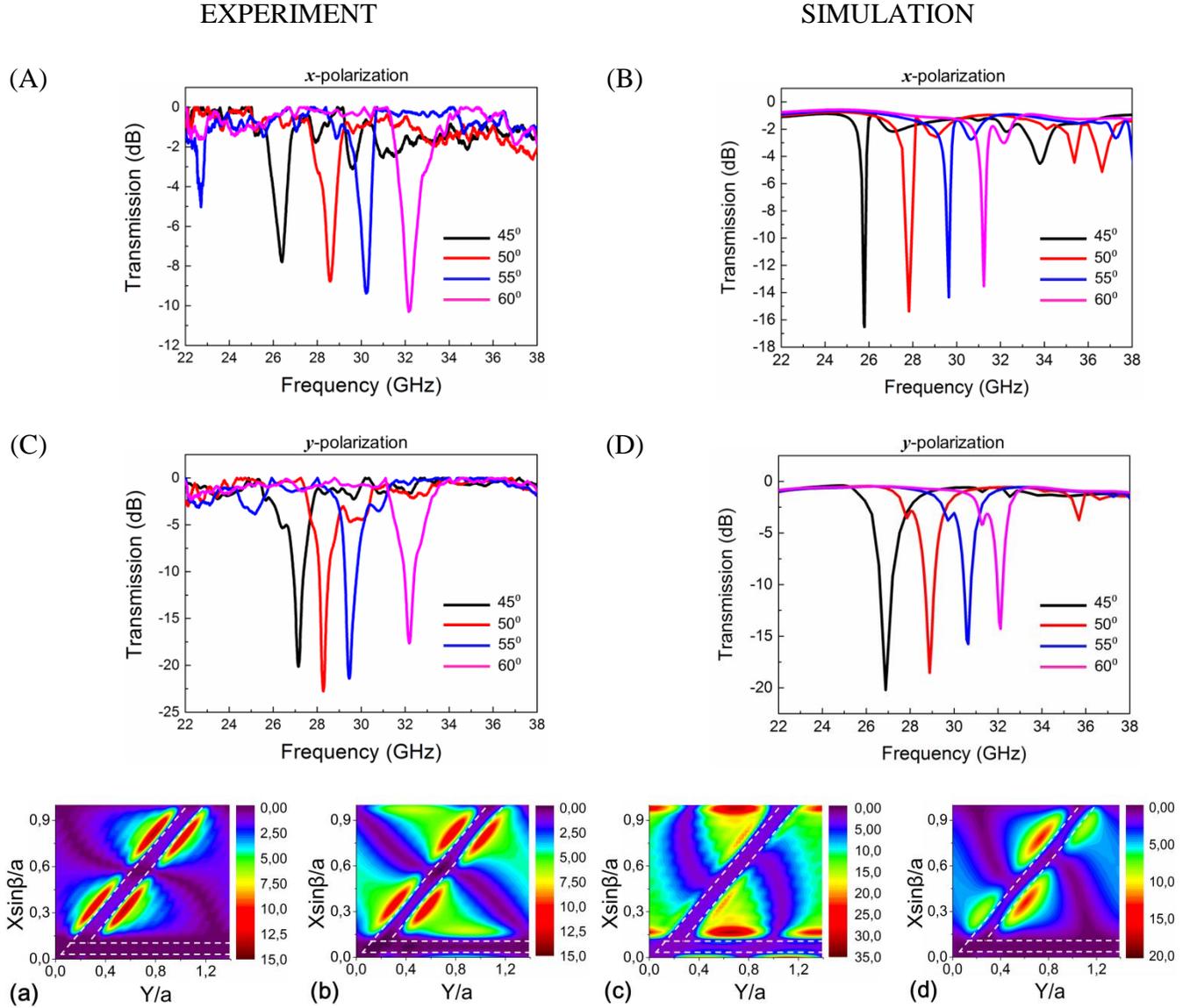

FIG. 3 (Color Online). Measured (left panel) and simulated (right panel) at normal incidence transmission spectra for gratings crossed at various angles: (a) calculated $|E_y|$-component of the instantaneous **E** field distribution on the shadow side of the unit cell at the resonance frequency for $\beta = 45°$, in the *x-y* plane at the distance *0.001a from metal strips*, where the normally incident electric field $\mathbf{E_0}$ is linearly polarized in the *y*-direction; (b) the same as in Fig.3 (a) for $|E_x|$-component; (c) the same as in Fig. 3 (a) for $|E_y|$-component however for the *x*-polarized incident wave; (d) the same as in Fig.3 (c) for $|E_x|$-component. White dotted lines indicate the boundaries of the metal strips

The integral equations are solved numerically using the integral functionals related to the polarization current distributions and the technique of the double Floquet-Fourier series expansions. In the final stage of the solution we obtain scattered fields as a superposition of the space harmonics in the forms given as follows

$$\mathbf{E}^r(x,y,z) = \sum_{p=-N}^{N}\sum_{q=-N}^{N} \mathbf{E}^r_{pq} e^{i(\xi_p x + \gamma_{pq} y - \kappa_{pq} z)} \quad (1),$$

$$\mathbf{E}^t(x,y,z) = \sum_{p=-N}^{N}\sum_{q=-N}^{N} \mathbf{E}^t_{pq} e^{i(\xi_p x + \gamma_{pq} y + \kappa_{pq} z)} \quad (2),$$

where

$\xi_p = k_x + \frac{2\pi p}{L_{x_1}}$, $\gamma_{pq} = k_y + \frac{2\pi}{\sin\beta}\left(\frac{q}{L_{y_1}} - \frac{p}{L_{x_1}}\cos\beta\right)$, $\kappa_{pq} = \sqrt{k^2 - \xi_p^2 - \gamma_{pq}^2}$ are the wave vector components of a diffracted order $p$ and $q$, $\mathbf{E}^r_{pq}$, and $\mathbf{E}^t_{pq}$ are the magnitudes of the electric field of the reflected and transmitted waves. For a crossed grating, the structure is periodic along two directions with periods $L_{x_1}$ and $L_{y_1}$ as illustrated in Fig.1 and the transformation between two coordinate systems $(x_1, y_1)$ and $(x, y)$ are $x_1 = x - y\cot\beta$, $y_1 = y\csc\beta$. The diffraction orders $p$ and $q$ are defined as harmonics propagating along grating vectors $X_1$ and $Y_1$, respectively. In our calculations we consider the periodic metallic grating as a perfect electric conductor (PEC) in the approximation given in [14, 18]. It is known that the fields scattered by a *PEC* object can be accurately modeled in terms of those scattered by a fictitious magneto-dielectric object satisfying the material conditions $|\varepsilon|\gg 1$ and $\varepsilon\mu = 1$. This approximation is effectively used in the analysis of the electromagnetic scattering by the gratings of PEC-like. Then we have assumed that $\varepsilon = 2.5 - i\,2500$ and $\mu = 1/\varepsilon$ to realize the material parameters for *PEC* material. For transmission theoretical calculations show a reasonable agreement with the experimental results (see Fig. 3 (A)-(D)) taking into account some imperfection of the measurement technique.

The typical simulated field distributions at resonance are shown in Fig. 3 (a)-(d). At resonant frequency, the incident electric field induces a dipolar oscillation mainly along the skew stripes, which has parallel $|E_y|$ and perpendicular $|E_x|$ components relative to $\mathbf{E}_0$ (see Fig. 3 (a), (b)), respectively. It can be seen that the field is practically not excited along the strips parallel to the direction of the incident field. Contrary to *y*-polarized incident plane wave (see Fig. 3(c), (d)) where at the resonance the induced field distributed around the perimeter of the periodic cell similar to the resonance modes of the whispering gallery or similar to the ring-like resonator modes. In this case, due to the greater interaction length i.e. when whole perimeter of the cell is the resonating portion so the field is stronger than that in the Fig. 3 (a), (b) where there is only side of the parallelogram is resonant part and in the first case as a consequence is higher quality factor for the resonance.

The resonant frequency *exhibits* almost linear dependence on the crossing angle of the strips of the gratings (see Fig. 4), where all resonances shift towards lower frequencies when the angle between the strips of crossed gratings decreases. Therefore, the curves demonstrate high degree of mechanical tunability of the metamaterial resonant response over a wide frequency range.

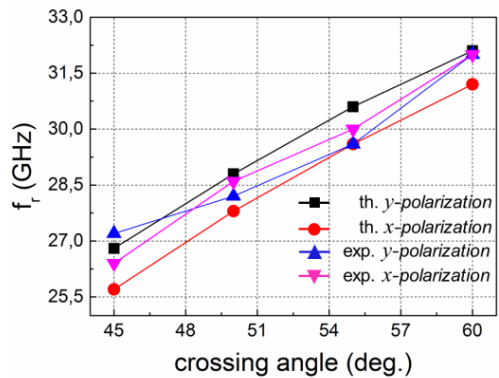

FIG. 4. (Color online) Theoretical and experimental resonance frequency as a function of crossing angle of the stripe gratings

In summary, mechanically tunable planar metamaterial named *metasurface with an oblique lattice* has been fabricated and investigated in the microwave part of spectrum. Utilizing two slabs with deposited metallic strip gratings a continuous tunability of the resonant frequency has been achieved by altering the angle between sandwiched crossed gratings through a wide frequency range. The metamaterials' response has a quality factor between 40 to 60 that higher than the typical value for many conventional metamaterials based on a split ring resonator. These results open the avenue for new design approaches of metamaterials that are mechanically tunable *metasurface with an oblique lattice*.